# Journal Name

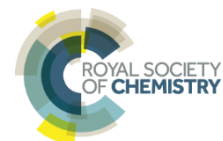

## ARTICLE

# Charge carrier transport and lifetimes in n-type and p-type phosphorene as 2D device active materials: an ab initio study

E. Tea,[a] C. Hin,[a,b]

In this work, we provide a detailed analysis of phosphorene performance as n-type and p-type active materials. The study is based on first principles calculation of phosphorene electronic structure, and resulting electron and hole scattering rates and lifetimes. Emphasis is put on extreme regimes commonly found in semiconductor devices, i.e. high electric fields and heavy doping, where impact ionization and Auger recombination can occur. We found that electron-initiated impact ionization is weaker than the hole-initiated process, when compared to carrier-phonon interaction rates, suggesting resilience to impact ionization initiated breakdown. Moreover, calculated minority electron lifetimes are limited by radiative recombination only, not by Auger processes, suggesting that phosphorene could achieve good quantum efficiencies in optoelectronic devices. The provided scattering rates and lifetimes are critical input data for the modeling and understanding of phosphorene-based device physics.

## Introduction

Since the first synthesis of graphene, more and more 2D materials are being discovered. These materials hold the promise of new lightweight and flexible electronic and optoelectronic nanodevices.[1-2] Phosphorene, single layer black-phosphorous, is the latest entry in the 2D material family tree with the recent successful synthesis of few- and mono-layer flakes by physical or chemical exfoliation from black-phosphorous.[3-4] Due to its puckered layer structure, phosphorene exhibits intriguing properties such as highly anisotropic electrical and thermal conductivities,[4-9] polarization dependent optical response,[10-13] and even a preferential direction for surface Li diffusion.[14] Its domain of application covers a very wide range, including transistors,[4,15-18] THz detectors,[19] solar cells and light emitting devices,[5,20] thermoelectric converters,[21,22] Li ion battery electrode,[14] and water splitting photo-catalyst.[23]

Phosphorene is characterized by a direct band gap, as well as small charge carrier effective masses. These features translate into an excellent $10^5$ current modulation ratio and charge carrier mobility up to 1000 cm$^2$/Vs in field effect transistor prototypes at room temperature.[4,15-18] These unique features make phosphorene an ideal candidate for future electronic and optoelectronic semiconductor nanodevices. This new 2D material improves on graphene, which exhibits a poorer current modulation due to its lack of band gap. Alternative materials, such as 2D semiconducting transition metal dichalcogenides, also have moderate band gaps but their carrier mobilities are 5-10 times lower than phosphorene.[24] Moreover, varying the number of phosphorene layers as well as stacking order modifies the band gap, making this material highly tunable.[11-12,25] While undoped phospherene exhibits p-type conductivity, ambipolar transport has been evidenced in phosphorene-based devices.[17-18,20] This ambipolar character can reduce the carrier conductivity affecting the entire 2D device performance. The tendency of phosphorene to oxidize easily can lead to the alteration of its mechanical and electronic properties that would ultimately degrade device performance.[26] This has been illustrated on electronic devices where encapsulation has been shown to preserve phosphorene structure and hence its electronic properties over time.[27] Thus the development of phosphorene-based devices strongly relies on the understanding of its charge carrier transport efficiency and recombination dynamics. Recent theoretical studies mainly focused on electronic structure calculation and associated absorption spectra,[10-13] electron mobility calculation,[17,28] extrinsic doping ability,[25] or thermal conductivity.[29-31] However, more advanced electronic transport properties beyond the sole electron-phonon and electron-impurity interactions, such as inelastic electron and hole scattering rates and recombination lifetimes, are required to understand and predict future device prospects and limitations.

In this account, phosorene is explored as n-type and p-type doped active materials under extreme regimes, i.e. high electric field and heavy doping, which are typically met in semiconductor devices. Under such regimes, hot carrier impact ionization and Auger recombination processes can take place. These processes are characterized by the non radiative generation and recombination of electron-hole pairs, whence

[a.] Department of Mechanical Engineering, Virginia Polytechnic Institute and State University, 635 Prices Fork Road, Blacksburg, VA, 24060, USA.
[b.] Department of Materials Science and Engineering, Virginia Polytechnic Institute and State University, 635 Prices Fork Road, Blacksburg, VA, 24060, USA.

their importance in semiconductor device physics. Hot electron gate injection followed by impact ionization in the channel has been linked to field effect transistor breakdown, strongly reducing device lifetime and reliability.[32-33] Auger recombination is known to limit minority carrier lifetimes, especially in narrow band gap semiconductors.[34-35] This strongly alters the efficiency of optoelectronic devices such as light emitting diodes where radiative recombination is crucial. Thus, both processes need to be explored since they dictate the material performance under different device operating conditions.

In the following, we formulate the numerical tools used for the calculation of the impact ionization and carrier-phonon scattering rates, as well as the radiative and Auger recombination lifetimes in single layer phosphorene. The input for such calculations is the electronic structure determined from first principles. Then, electrons and holes scattering rates, lifetimes and their impact on material properties are presented in the light of the electronic structure for a wide range of doping levels and temperatures for the carrier distributions. Finally, these properties are discussed in various device contexts.

## Numerical Tools

Scattering rates and lifetimes have been calculated from the Fermi's Golden Rule:

$$S_{k,k'} = \frac{2\pi}{\hbar} |M_{k,k'}|^2 \delta(E)$$

where $S_{k,k'}$ gives the probability per unit time of a transition between a given pair of initial and final states labeled by their wave vector $k$ and $k'$, respectively. $M_{k,k'}$ is the interaction Hamiltonian evaluated between these states. $\delta(E)$ implies the overall conservation of energy across the transition. The impact ionization and Auger recombination processes involve two charge carriers. Their initial and final states $k$ and $k'$ are composite states $(k_1, k_2)$ and $(k'_1, k'_2)$. The total scattering rate $w_{k1}$ for an initiating carrier $k_1$ defined as a conduction electron, is given by a threefold summation of $S_{k,k'}$ over all partner carriers $k_2$, and all allowed final states $(k'_1, k'_2)$ fulfilling energy and momentum conservation. Umklapp processes are accounted for, using reciprocal space periodic boundary conditions.

The input for the Fermi's Golden Rule is the electronic structure of phosphorene. It provides (i) the energy-wave vector relation required to evaluate $\delta(E)$, and (ii) the associated wave functions required to calculate the interaction strength $M_{k,k'}$. The electronic structure of single layer phosphorene has been calculated using Density Functional Theory, as implemented in the Vienna Ab initio Simulation Package within the projector augmented wave framework.[36] To correct the semiconductor band gap underestimation tendency of conventional exchange and correlation functionals, such as LDA or GGA, the PBE0 hybrid exchange and correlation functional has been used.[37-38]

To discuss the overall effect of impact ionization and Auger recombination on transport properties in phosphorene, other processes such as carrier-phonon scattering and radiative recombination also need to be accounted for. Radiative recombination lifetimes are calculated directly from the electronic structure. The photon polarization is chosen parallel to the armchair direction since it exhibits the strongest electron-photon coupling, and consequently the shortest radiative lifetime possible. Other groups have also evidenced this strong anisotropic optical coupling in monolayer[10] and bi- or tri-layer phosphorene irrespective of the stacking order.[11] At room temperature carrier-phonon scattering dominates the charge carrier transport, over carrier-impurity scattering,[15] and will then serves as a basis for comparison with impact ionization. The determination of carrier-phonon scattering rates requires the phonon dispersion relation. The Einstein and Debye models have been used for the optical and acoustical phonon dispersions, respectively. As presented in Ref. 39, the flexural mode is neglected because it would largely be suppressed in a non-freestanding phosphorene flake. The optical phonon energy of 0.055 eV has been estimated from density functional perturbation theory and molecular dynamics calculations presented in Refs. 21 and 31. Direction dependent acoustic phonon velocities are taken from Ref. 9. Phonon populations are set to room temperature. In our study, carrier-phonon deformation potentials are assumed to be constant and have been determined by fitting the scattering rates to those of Liao et al., which involves more detailed calculations close to the band edges.[39]

## Results

### 1- Electronic structure

The calculated lattice parameters of 4.54 Å and 3.27 Å along the armchair and zigzag directions agree well with previous computational reports on phosphorene.[10-11,13] The calculated 2.2 eV electrical band gap is also in agreement with other calculations, notably using the GW approach to calculate quasi-particle energy levels.[11-12,39] Large exciton binding energies (0.7 - 0.8 eV) have been calculated in phosphorene by solving the Bethe Salpeter Equation.[11-12] These results are supported by the 1.45 eV optical band gap measured by photoluminescence,[4] and the 2.05 eV electrical band gap measured by scanning tunneling spectroscopy.[40] Hence, our band structure calculated with the PBE0 hybrid exchange and correlation functional is relevant to discuss the electronic properties of phosphorene. The conduction band minimum (CBM) and valence band maximum (VBM) are both located at the Γ point. In Ref. 39, the conduction and valence bands of phosphorene exhibit a pronounced anisotropy, as shown in our Figure 1 as well. Assuming a parabolic dispersion model, effective masses were extracted from our band structure and are collected in Table I. In the armchair direction, the effective masses are about 5 times smaller than in the zigzag direction, which indicates a preferential transport along this direction for both electrons and holes under low field conditions. Besides the CBM, the conduction band exhibits a local minimum in the zigzag direction. This satellite valley is located only 0.2 eV above the CBM as indicated in Figure 1. This suggests that a single valley model for the conduction band would not be adapted to the description of electron transport. Indeed, this valley is close enough to the CBM energy that electrons could populate it under moderate electric fields or temperature regimes.

|  | armchair | zigzag |
|---|---|---|
| VBM | 0.18 | 8.38 |
| CBM | 0.20 | 1.02 |
| local min. | 0.22 | 0.32 |

Table I: Direction dependent effective masses in units of the free electron mass, for the valence band maximum, the conduction band minimum, and the local minimum shown in Fig. 1.

## 2- Impact ionization and carrier-phonon scattering

The calculated impact ionization scattering rates as a function of the initiating carrier energy is shown on Figure 2a and 2b, for conduction electrons and valence holes respectively. The zeros of the energy axis have been set to the CBM and the VBM. The rates given as a function of energy are obtained as the sum of the momentum dependent rates $w_{k1}$, weighted by the initiating carrier density of states. Indeed, impact ionization is a high field process involving a hot impacting electron with a kinetic energy larger than the band gap energy. At such high energies the anisotropy observed close to the band edges is effectively smoothed out due to carrier-phonon scattering. Hence, while being anisotropic in the Brillouin Zone, energy averaged rates provide relevant pictures of electron scattering. The impact ionization scattering rates steeply increase beyond the threshold energy, indicating a large increase of the density of final states, for both electron and hole initiated processes. The threshold energy can be extracted by fitting the scattering rates to the Keldysh expression $A \times ((E - E_{th})/E_{th})^B$ (see Figure 2 caption). The calculated threshold energy $E_{th}$ of 2.28 eV is found to be very close to the band gap energy. This indicates that impact ionization processes could easily take place in phosphorene, without phonon-assisted processes. However, close to the threshold energy, the impact ionization scattering rates seem fairly low when compared to the carrier-phonon ones. Our calculations show that carrier-phonon interaction is generally stronger than impact ionization. This holds for electrons with energies up to 5 eV, and holes with energies up to 4 eV in excess of the band edges. These energies are very high, and correspond to about twice the impact ionization threshold energy. Thus, impact ionization processes are not expected to become dominant over carrier-phonon processes under moderate electric field conditions.

## 3- Radiative and Auger lifetimes

Minority carrier lifetimes have been calculated and are shown in Figure 3 for p-type (minority electrons) and n-type (minority holes) phosphorene. Over the investigated doping level range (from $1 \times 10^{11}$ to $1 \times 10^{14}$ cm$^{-2}$), minority electrons lifetimes are limited by radiative recombination only, while minority holes go from a radiatively limited, to an Auger limited regime for doping densities beyond $1 \times 10^{12}$ cm$^{-2}$. This suggests that Auger processes where the recombination energy is given to an electron (eeh processes) are more efficient at recombining electron-hole pairs than processes where the energy is given to a hole (ehh processes). Minority electron radiative lifetimes reach a plateau for doping densities beyond $2 \times 10^{13}$ cm$^{-2}$. Indeed, increasing hole density leads to a broader occupation of reciprocal space by holes due to the Pauli Exclusion Principle and the very small valence band curvature in the zigzag direction. Meanwhile, the conduction band is characterized by larger band curvatures, limiting the spreading of the electron distribution when increasing the electron density. Consequently, the hole density increase leads to a saturation of the number of direct electron-hole transition channels available for radiative recombination. This saturation phenomenon also takes place for minority holes, and is already reached for majority electron densities as low as $1 \times 10^{11}$ cm$^{-2}$. The saturation threshold is lower in the case of minority holes for the same reasons.

Assuming a frozen band structure, the minority carrier lifetimes have been calculated for carrier distributions between 300 and 600K. Increasing the charge carrier temperatures tend to shorten lifetimes, since heated carrier distributions can access larger reciprocal space regions using the extra thermal energy. This provides more channels to the heated carriers to fulfill energy and momentum conservation in radiative and Auger recombination processes. However, the temperature increase has no effect on the Auger limited minority hole lifetimes for very large electron densities. Heavily degenerated electron populations already occupy a large region of reciprocal space because of the Pauli Exclusion Principle, making the temperature effect anecdotic. This is illustrated on Figure 4, where electron distributions in reciprocal space are given for different values of electron density and temperature. Figure 4 also shows that increasing the electron density, or the temperature, lead to the occupation of the satellite valley as pointed out on Figure 1 in the zigzag direction. Occupation of this satellite valley gives electrons access to a larger region of reciprocal space, giving them more channels to fulfill energy and momentum conservation in Auger recombination, and may explain the strength of eeh processes.

In the case of electron-hole pairs created by photo excitation, additional calculations have been performed with a 1.7 eV band gap, smaller than the electrical band gap, by manually shifting the bands to simulate exciton binding energies. Similar lifetimes have been obtained, differing by much less than one order of magnitude.

## Discussion

### 1- Electronic devices

In field effect transistors, hot electron injection through the gate dielectric can be followed by impact ionization in the channel. This can lead to local heating, the creation of defects, and ultimately to device failure.[32-33] To build ultrathin devices, it has been proposed to use phosphorene as the active material in conjunction with other 2D van der Walls materials such as hexagonal boron nitride (h-BN).[10] Successful prototype devices with h-BN as dielectric layers have been reported.[16,20] It has been observed that h-BN makes cleaner semiconductor-insulator interfaces compared to commonly used $SiO_2$ dielectrics, and is more resilient to soft dielectric breakdown.[41-44] Moreover, h-BN exhibits a large 6 eV band gap, and a high breakdown field (12 MV/cm) comparable to that of conventional $SiO_2$ dielectrics.[44] h-BN is then expected to perform better than $SiO_2$ because it would also protect phosphorene from oxidation and preserve its structure and electronic properties.[27] h-BN is considered as the gate dielectric in this discussion. Large gate bias can cause electron tunneling from the gate electrode, and then lead to the injection of electrons from the h-BN conduction band to the phosphorene layer. This will give birth to hot carriers in phosphorene that can trigger impact ionization. The band alignment between h-BN and phosphorene is difficult to deduce from Ref. 10 due to the large smearing used in the calculation of the projected density of states. However, a 2.5 eV barrier can be estimated from electron affinities difference. Then, 2.5 eV would also be a lower estimate of the electron excess energy after hot carrier injection. Figure 2 showed that impact ionization processes would compete with carrier-phonon scattering for hot carriers with excess energies close to or beyond 4 eV. Under very thin dielectric thickness and very large gate bias conditions, such electron excess energies are still possible. This indicates that, although h-BN/phosphorene interfaces seem very promising, hot carrier injection can become problematic for devices with single layer phosphorene/h-BN interfaces under very large electric fields. Multi-layer phosphorene exhibit narrower band gaps than the single layer,[12] which would lower their impact ionization threshold energies. Hence, hot carrier injection followed by impact ionization appears to be even more critical for multi-layer phosphorene/h-BN interfaces. In a regime of heavy impact ionization, the extent to which Auger recombination can suppress the newly created electron-hole pairs is not clear, and more details on the charge carrier dynamics is required. However, Figure 3 shows that electron impact ionization would be more affected by recombinations than hole impact ionization because of the shorter recombination lifetimes of the n-type case. In opposition, recombination lifetimes are longer in the p-type case. Consequently, possible hole initiated impact ionization cannot be efficiently compensated by Auger recombination, and can be detrimental to p-type device reliability.

**2- Optoelectronic devices**
Phosphorene has also been proposed as the active material in optoelectronic devices. Its direct band gap and ambipolar transport properties make it an excellent candidate for homojunction optoelectronic devices.[15,17] Moreover, pn-homojunctions showing a photovoltaic effect have been experimentally demonstrated.[20] To make successful optical emitters or detectors, non-radiative recombination should be avoided. Figure 3 shows that in p-type phosphorene, charge carrier lifetimes are limited by radiative recombination only, and Auger recombination is not likely to occur. Consequently, phosphorene is a very strong p-type material candidate for optoelectronic devices. Used as an n-type material, phosphorene can still show good optoelectronic properties, although in a limited electron density range. The radiative regime is maintained up to an electron density of $1\times10^{12}$ $cm^{-2}$ with lifetimes longer than 10 ns. The recombination lifetimes become strongly limited by Auger recombination for electron densities beyond $1\times10^{12}$ $cm^{-2}$. This would largely impair n-type phosphorene optoelectronic properties for heavy electron doping. However, a phosphorene pn-homojunction with tight control of the different regions width could lead to LEDs with excellent quantum efficiencies.

**3- The satellite valley**
In phosphorene, p-type transport has already been praised for its better mobilities and conductivities compared to n-type transport.[4,15] In addition, p-type phosphorene has the ability to form ohmic contacts with usual metals such as titanium or gold,[15] without having to resort to graphene contacts, which is a technological asset. The poorer n-type transport properties can be explained by (i) larger interaction rates with phonons (compare Figure 2a and 2b), and (ii) the presence of a satellite valley only 0.2 eV above the CBM (see Figure 4). Indeed, the presence of an extra valley close to the conduction band minimum has several implications, as seen in GaAsSb for example.[45] First, electrons in the satellite valley virtually have less kinetic energy than in the main valley, because of the 0.2 eV offset. This translates into overall smaller electron velocities than if the material only had a single conduction valley. Second, the satellite valley increases the density of states, and gives birth to intervalley electron transfer. Intervalley scattering randomizes electron momentum, ultimately slowing down the electron population. However, n-type transport can be improved by stress engineering, by modifying effective masses and intervalley energy offsets.[7] One can also take advantage of the satellite valley to open new degrees of freedom in optimizing materials for thermoelectric applications. Indeed, the conduction band anisotropy can combine a large density of states with high electron mobility. Engineering the density of states via the satellite valley would provide an additional control between doping density and the Fermi level, which can improve the Seebeck coefficient.[39]

## Conclusions

Electron and hole scattering rates and lifetimes in single layer phosphorene have been calculated *ab initio* alongside the electronic structure. Their role and impact under high electric field and heavy doping conditions have been discussed in various device contexts. The very sensitive link between electronic structure, carrier distribution, scattering rates and lifetimes has been evidenced. It explains the poor n-type conductivity, and predicts that Auger recombination will not be an issue in p-type. To further understand the prospects and limitations of phosphorene as an active material in a nanodevice as a function of electrode metals and substrate or encapsulating materials, device level modeling of charge carrier transport is required. Such modeling can be performed using the Ensemble Monte Carlo method, using our results as building blocks. Accounting for various encapsulating materials and the resulting stress conditions could open new degrees of freedom in optimizing the relation between doping and Fermi levels via the satellite valley. These are of interest for thermoelectric applications, in addition to the anisotropic electronic and thermal conductivity.

## Acknowledgements

This work was funded by the Air Force with program name: Aerospace Materials for Extreme Environment, and grant number: FA9550-14-1-0157. We acknowledge Advanced Research Computing at Virginia Tech for providing the necessary computational resources and technical support that have contributed to this work (http://www.arc.vt.edu).

## Notes and references

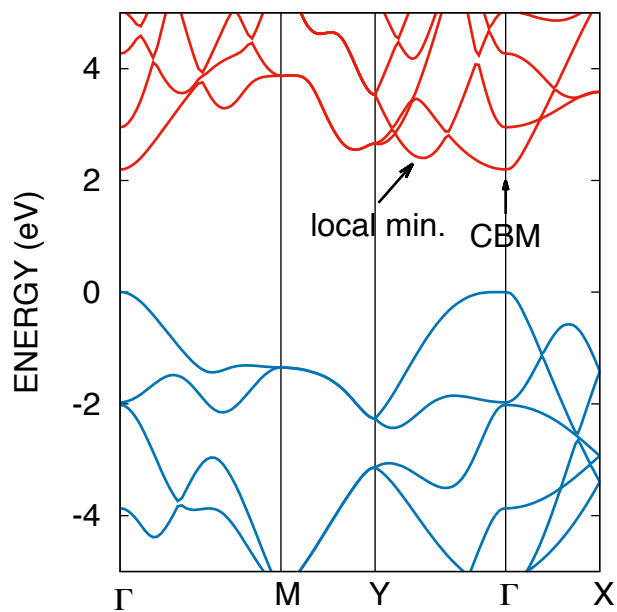

**Figure 1.** Calculated electronic structure of phosphorene. The band gap is direct at Γ. Anisotropy around the band edges arises between the armchair (Γ-X) and zigzag (Γ-Y) directions. A local minimum lies only 0.2 eV above the conduction band minimum (CBM) in the zigzag direction.

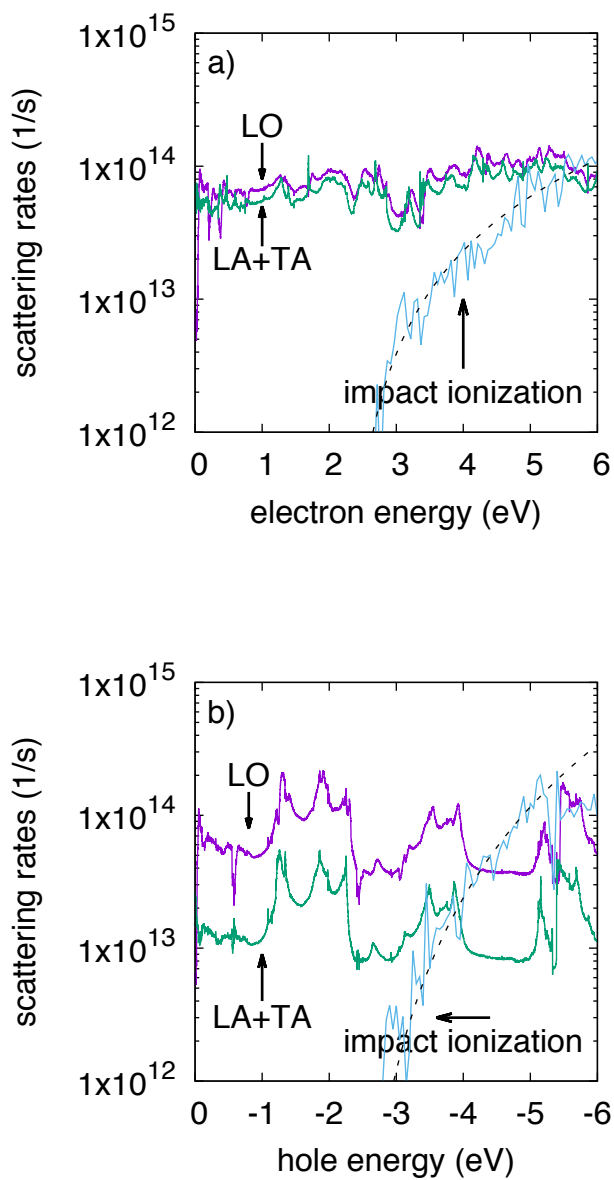

**Figure 2.** Impact ionization scattering rates as a function of a) impacting electron energy, and b) impacting hole energy. Electron-phonon and hole-phonon scattering rates are shown for comparison. The impact ionization rates can be adjusted by $A \times ((E - E_{th})/E_{th})^B$ $A \times \left(\frac{E-E_{th}}{E_{th}}\right)^B$ with $A = 4.12 \times 10^{13}$, $B = 2.05$ for electrons, and with $A = 6.30 \times 10^{13}$, $B = 3.44$ for holes (dashed lines). The threshold energy is $E_{th} = 2.28$ eV in both case.

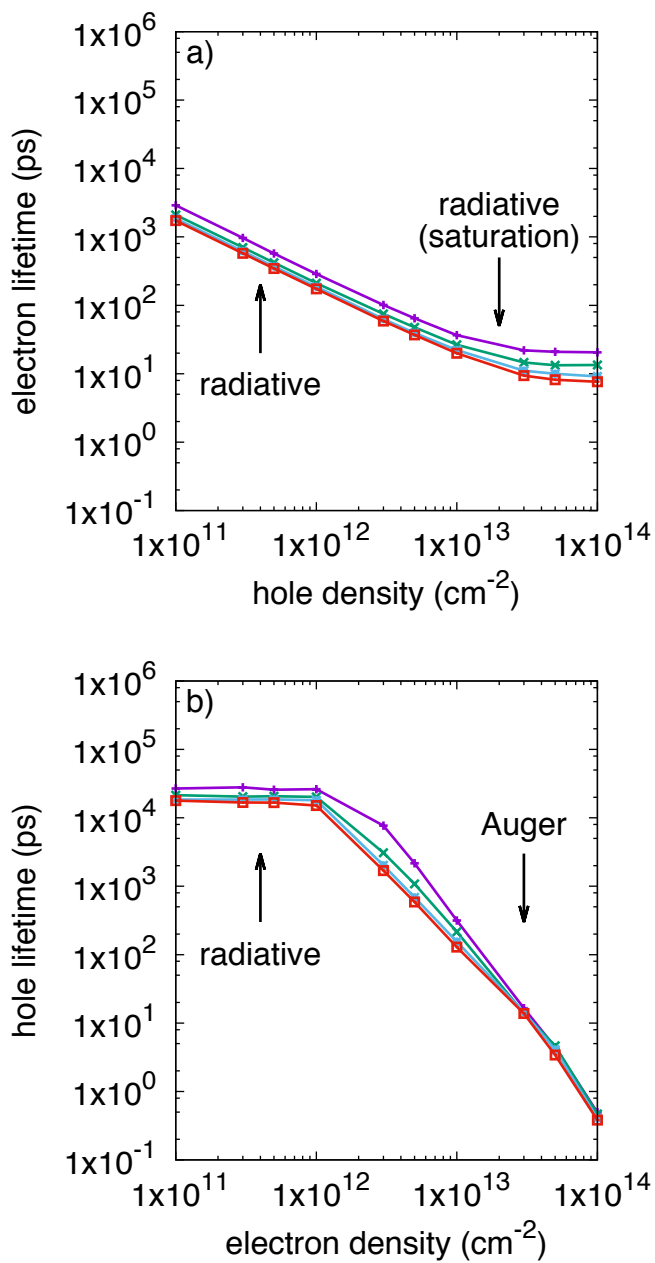

**Figure 3.** Calculated recombination lifetimes of a) minority electrons, and b) minority holes as a function of doping density. Lifetimes for heated carrier populations at 300, 400, 500 and 600K are shown in purple (+), green (x), blue (square) and red (open square) respectively.

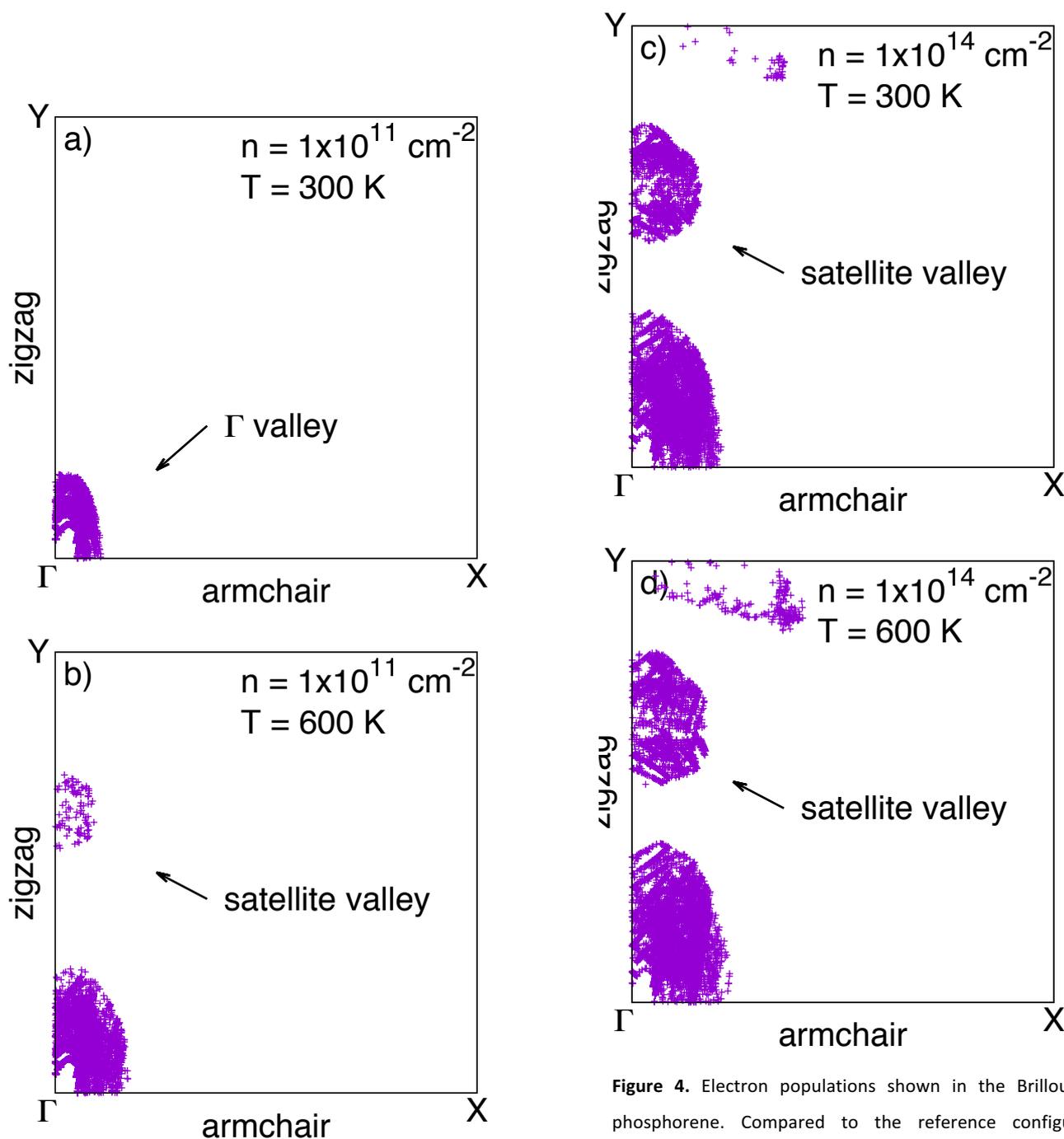

**Figure 4.** Electron populations shown in the Brillouin Zone of phosphorene. Compared to the reference configuration (a), increasing the distribution temperature (b) or the electron density (c) leads to the population of the satellite valley in the zigzag direction.